# Securing Mobile Ad hoc Networks: Key Management and Routing

Kamal Kumar Chauhan[1] and Amit Kumar Singh Sanger[2]

[1]Department of IT, Anand Engineering College, Agra, India

kamalchauhans@gmail.com

[2]Department of CSE, Hindustan College of Science and Technology, Mathura, India

sanger.amit@gmail.com

## ABSTRACT

*Secure communication between two nodes in a network depends on reliable key management systems that generate and distribute keys between communicating nodes and a secure routing protocol that establishes a route between them. But due to lack of central server and infrastructure in Mobile Ad hoc Networks (MANETs), this is major problem to manage the keys in the network. Dynamically changes in network's topology causes weak trust relationship among the nodes in the network. . In MANETs a mobile node operates as not only end terminal but also as an intermediate router. Therefore, a multi-hop scenario occurs for communication in MANETs; where there may be one or more malicious nodes in between source and destination.  A routing protocol is said to be secure that detects the detrimental effects of malicious node(s in the path from source to destination). In this paper, we proposed a key management scheme and a secure routing protocol that secures on demand routing protocol such as DSR and AODV. We assume that MANETs is divided into groups having a group leader in each group. Group leader has responsibility of key management in its group. Proposed key management scheme is a decentralized scheme that does not require any Trusted Third Party (TTP) for key management. In proposed key management system, both a new node and group leader authenticates each other mutually before joining the network. While proposed secure routing protocol allows both communicating parties as well as intermediate nodes to authenticate other nodes and maintains message integrity.*

## 1. INRODUCTION

A MANET is a special type of wireless network in which mobile hosts are connected by wireless interfaces forming a temporary network without any fixed infrastructure. In MANET, nodes communicate each other by forming a multi-hop radio network. Mobile nodes operate as not only end terminal but also as an intermediate router. Data packets sent by a source node can reach to destination node via a number of hops. Thus multi-hop scenario occurs in communication and success of communication depends on nodes' cooperation.

DOI : 10.5121/ijans.2012.2207    65

Security of a network is an important factor that must be considered in constructing the network. A network has to achieve security requirements in terms of authentication, confidentiality, integrity, availability and non repudiation. These security requirements rely on the availability of secure key management system in network. Fundamental goal of a key management system in a network is to issue the keys to the nodes to encrypt/decrypt the messages, to manage these keys and to prevent the improper use of legally issued keys. Absence of key management system makes a network vulnerable to several attacks [6]. Therefore, key management system is the basic and important need of a network for secure communication. A key management system normally involves key generation, distribution, updation and revocation of keys in network. The feature of MANETs such as dynamic topology, lack of centralized authority, resource constrained and node mobility are the major challenge in establishment of key management. Some techniques such as intrusion detection mechanism consume lot of nodes' battery power but cannot account for flexible membership changes. However, an efficient and secure key management system can solve this problem with an affordable cost.

On the hand, mobile ad hoc networking is multi-hop relaying, i.e. messages are forwarded by several mobile nodes from source to destination, if destination node is not directly reachable. In other words, nodes in MANET operate as not only end terminal but also as an intermediate router. Thus, multi-hop scenario occurs; where an attacker can insert, intercept or modify the messages easily in absence of secure routing protocol. This means that unprotected MANET is vulnerable to many attacks [21] such as wormhole attack [22], black hole attack [23] including node impersonation, message injection, loss of confidentiality etc.

In this paper, we proposed a key management scheme for group based MANETs in which a group leader can generate, distribute, update and revoke keys in its group and a provable secure routing protocol. Proposed key management scheme neither depends on a central server nor is it fully distributed. Our key management system forms a decentralized system that combines both centralized key management as well as distributed key management so that it can combine merits of both methods. Proposed key management scheme is a hybrid key management scheme that uses both Symmetric Key Cryptography (SKC) for secure communication and Public Key Cryptography (PKC) to authenticate other nodes and to share a session key.

We also proposed a secure routing protocol especially for On-demand routing protocol. Objective of proposed routing protocol is to authenticate the source and destination and intermediate nodes in route list of route request (RREQ) message and detecting any kind of modification by a malicious node in RREQ message, providing secure route. Proposed protocol also allows to intermediate nodes to authenticate its predecessor node, and then rebroadcast the RREQ message. Finally at destination, all nodes are authenticated and checked message integrity and then sends back route reply (RREP) message towards source.

Rest of the paper is organized as follows. In section 2, grouping and algorithm to elect group leader is discussed. Key management system is proposed in section 3. In section 4, we proposed provable secure routing protocol. Security analysis of proposed key management system and routing protocol is discussed in section 5 and finally, section 6 gives conclusions.

## 2. GROUP FORMATION

Grouping or clustering is a process that divides the network into interconnected substructure known as groups. Grouping provides a better solution to the problem of key management and routing in MANET. There is a group leader as coordinator in every group. Each group leader acts as a temporary base station within its zone or group and communicates with other group leader.
A system model of open MANET is shown in Figure.1. Mobile nodes are divided into several groups in such a way that all the nodes are covered with no groups overlapped. Some of the nodes are selected as group leaders to perform the functions of key management system and other administrative functions in its group. Aim of constructing the grouped based structure is that grouping preserves the structure of network as long as possible, when nodes moves or topology is slowly changing. On the other hand, grouping reduces the number of keys, required to distribute in network for secure communication.

Group based structure distributes the functions of a central server into several nodes (group leaders). Therefore, it combines both centralized and distributed approaches of key management system providing a decentralized solution. Group based structure of networks also removes the vulnerability of compromising single central server. If a group leader is compromised; only a group will be compromised leaving rest of the network safe and secure.

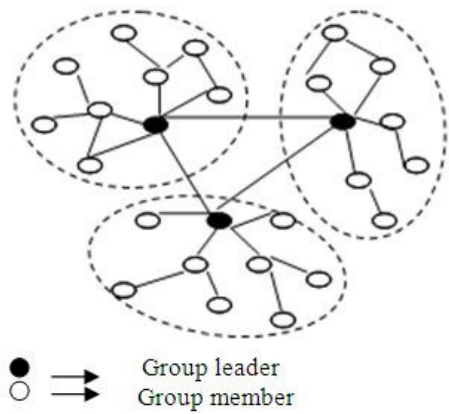

Figure 1. System model for MANET

## 2.1 Algorithm: Electing Group Leader in a Group

A good grouping algorithm is one that divides the network into groups in such a way that it preserves the structure of network as long as possible and fast recovery from fault such as electing new group leader on the failure of existing group leader.

To select a well suited group leader, we take into account of its mobility, battery power and behavior of node. The following features are considered for grouping:

- Each group leader is capable to support maximum 'x' number of nodes (a pre-defined value) efficiently. If a group leader is trying to serve more than 'x' nodes, system's efficiency suffers.
- 'Mobility' is the important factor in deciding the group leader. Group leaders are responsible to preserve the structure of group as much as possible when nodes move. Moving group leader quickly results detachment of nodes from group leader and also increases the probability of nodes' compromised. Mobility of a node is denoted by 'M' and can be measured as:

$$M = \frac{1}{T}\sum_{t=1}^{T} \sqrt{(X_t - X_{t-1})^2 + (Y_t - Y_{t-1})^2}$$

Where, (Xt, Yt) and (Xt-1, Yt-1) are the coordinates of a node at time t and t-1.

- 'Battery power' (B) is another important factor to decide a group leader. A group leader consumes more battery power than an ordinary node because a group leader has extra responsibilities such as monitoring group members and distribution of keys in the group. Therefore, node with maximum battery power should be elected as group leader.

- Another important parameter for electing the group leader is the 'behavior of node'. Security of a group is totally depends on group leader. Group Leader monitors the nodes' activities continuously in the group and assigns them a Trust Level (T) on the basis of their behavior.

Finally, group leader is selected on the basis of weight (W), is defined as:

$$W = w_0 M + w_1 B + w_2 T$$

where, $w_0$, $w_1$, and $w_2$ are the weight factor such as:

$$w_0 + w_1 + w_2 = 1$$

Select a node as group leader with the smallest weight (W).

## 3. PROPOSED KEY MANAGEMENT SCHEME

In this section, we proposed key management system for group based MANETs. Proposed key management scheme includes key generation, distribution and revocation phase. We make following assumptions:

- An offline Trusted Third Party (TTP) is available outside the network which is responsible only to issue a certificate and public/private key pair for mobile nodes.
- Intergroup communication is done through group leaders.
- Group leaders are trusted. Grouping algorithm is not periodic. This reduces updates and hence computation and communication cost in system.

### 3.1 Key Generation and Distribution

All group leaders in network are assigned a unique id. Each group leader has a public/private key pair and a secure hash function (for e.g. SHA or MD5). We define three types of keys in the network: Group key, key for all the members in group used to encrypt/decrypt all the traffic communicated in the group. Second key, a symmetric key shared between group leader and a member node of same group and third key, shared by all group leaders in network.

Group leaders generate group key for their groups independently. Group key is updated each time when a node joins or leaves the group to maintain the forward and backward secrecy. Second key (k) is shared between group leader and a member node at the time when node joins group. k is the function of node_id and a secret randomly generated number by group leader.

$$f(node\_id, N) = k$$

where f is a secure hash function selected by group leader, node_id is assigned to a node at the time of joining and N is a secret number known only to group leader.

Third key is shared by group leaders in network. Group leaders can agree on a key to communicate securely using Group Diffie-Hellman key agreement protocol [13]. Key is updated when group leader election algorithm is invoked in any group; new elected group leader can start Group Diffie-Hellman key agreement to update the key.

### 3.2 Node addition

Whenever, a new node joins a group. It sends a request to group leader. This request might be captured by a malicious node showing as group leader to new node. Similarly, a malicious node can also send a request to group leader to join the group. Therefore, it is necessary for both group leader as well as new node to authenticate each other. Upon successfully mutual authentication, a node can join the group and share a key with group leader in a secure manner. A new node and group leader can authenticate each other using challenge-response protocol. New node sends a challenge to group leader and group leader provide a valid response to prove its genuinity.

Group leader selects two large prime numbers 'p' and 'q' and calculates: N=p*q, then selects a random secret number 'S' and calculates: $V=S^2$ mod N (1<S<N).

'N' and 'V' are publically announced in the group. When group leader has to authenticate itself i.e. it received a challenge from a node, it finds $X=R^2$ mod N, where 'R' is a random number selected by group leader such that 1<R<N.

Group leader sends {N, V, X} to new node. On receiving (N, V, X), new node sends a challenge 'c' to group leader. Group leader calculates $Y=RS^C$ mod N and send it to node. Node calculates $XV^C$ and match with $Y^2$. If both values are same, group leader is successfully authenticated.

After successful authenticating to group leader, new node can sends its certificate to group leader issued by offline TTP. Group leader verifies nodes' certificate, and extracts the public key of node from certificate. Group leader generates a node_id and sends node_id and a key generated by function f shared by group leader and node, encrypted with public key of new node. Group leader then update group key and group members list and sends to the members of group. Communication between group leader and new node takes place as follows:

A group of mobile nodes with a group leader of MANET is shown in Figure.2, where a new node 'A' wants to join the group. Following are the notations used in communication:

G  ⟶  Group, {L, M}

M  ⟶  Set of group members {$m_1, m_2, m_3 \ldots m_n$}

L  ⟶  Group leader

A  ⟶  New node

$ID_A$  ⟶  A's Identity given by group leader

$K_{XY}$  ⟶  Session key shared between node X and Y

$e_X/d_X$  ⟶  Public key/Private key of node X

$DS_X$  ⟶  Digital Signature of node X

$T_X$  ⟶  Timestamp added by node X

$CERT_X$  ⟶  Certificate of node X

$S_{LX}$ ⟶  Symmetric key shared between group leader and node X.

X: Y {k(M)} ⟶ Node X sends a message M encrypted with key k to node Y

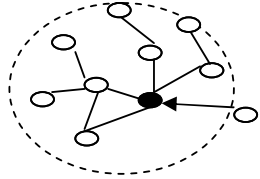

Figure 2. A Group in MANET

- A  :  L {A, Join_req}
- L  :  A ∪ M {N, V, X}
- A  :  L ∪ M {c}
- L  :  A ∪ M {Y}
- A  :  L {CERT$_A$}
- L  :  A {e$_A$ (e$_L$, ID$_A$, S$_{LA}$)}
- A  :  L {S$_{LA}$ (num}
- L  :  A {S$_{LA}$ (num, member_list, group key)}
- L  :  M {group_key (new_group_key)}

### 3.3 Key Agreement Protocol

If a node A wishes to communicate securely with node B. Before starting communication, they must agree on a session key. A starts communication by sending message:

A  : B {e$_B$ (ID$_A$, ID$_B$, T$_A$, DS$_A$)}

On receiving message from A, B decrypts the message and verifies the signature of A using public key of A. If node B does not have A's public key, it sends a message to group leader conveying to send A's public key. Here following two cases are possible:

- A is a genuine node and group leader has pubic key of A. In this case, group leader sends A's public key to B. B then verifies A's signature and share a session key $K_{AB}$.

    B  :  A {e$_A$ (ID$_A$, ID$_B$, T$_A$, T$_B$, DS$_B$)}

    A  :  B {e$_B$ (T$_A$, T$_B$, num1, K$_{AB}$)}

    B  :  A {K$_{AB}$ (num1, num2)}

- In second case, A is malicious node and not a member of group. In that case, group leader would inform to all the member of group about node A.

International Journal on AdHoc Networking Systems (IJANS) Vol. 2, No. 2, April 2012### 3.4 Node Deletion

Nodes in a group communicate with group leader periodically showing its presence in group. If a node doesn't communicate, group leader removes that node from member list and intimate other member. Group leader regenerates new group key and sends other nodes in group, encrypted by their public key. A node can be removed from member list when one of the following events occurs:

- A node can leave the group with prior notification.
- A node can leave the group without any prior notification or node is not forwarding the messages or performing as malicious node. Group leader exclude that node forcefully. In this case, group leader must inform to neighbor leader nodes.

On the other hand, whenever a group leader left the group with or without prior notification, a new group leader must be elected that can coordinate the group. New group leader reconstructs new group key and distributes in the group encrypted with the public key of members and share a new symmetric key with each member in group. New group leader distributes its public key and id to other group leader in network and starts Group Diffie-Hellman key agreement [13] to update key shared by group leaders.

## 4. PROVABLE SECURE ROUTING PROTOCOL

In this section we proposed a secure routing protocol in which source and intermediate nodes append their digital signature and hash code of received RREQ message to route request (RREQ) message and then rebroadcast RREQ message. When neighbors of source receive RREQ, they verify the signature of source and make the decision accordingly. Destination sequence number [25] in the protocol is added to make the loop free routing and to check the freshness of route control packet. Source and destination node may or may not be in the same group. We discussed both cases i.e. intra group communication and inter group communication.

### 4.1 Intra Group Communication

Assume that S is the source node trying to discover a route to destination D. A and B are two intermediate node. All nodes are in the same group.

**Notations:**

1) $L_{RREQ} \rightarrow$ Life time (maximum number of hops) of RREQ.
2) Seq $\rightarrow$ Destination sequence number.
3) $DS_A \rightarrow$ Digital Signature of node A.
4) $h_A \rightarrow$ Hash code appended by A to RREQ.
5) A $\rightarrow$ *   A broadcasts message.
6) A $\rightarrow$ B   A sends message to B.
7) H $\rightarrow$   Hash function

**Route Discovery:**

1) S $\rightarrow$ * RREQ (S, D, Seq, $L_{REQ}$, <S >, $DS_S$, $h_S$)
2) A $\rightarrow$ * RREQ ( S, D, Seq, $L_{REQ}$-1, <S, A>, $DS_S$, $DS_A$, $h_A$)
3) B $\rightarrow$ * RREQ (S, D, Seq, $L_{REQ}$-2, <S, A, B>, $DS_S$, $DS_A$, $DS_B$, $h_B$)

71

4) D → B    RREP    (S, D, Seq, <S, A, B, D>, DSs, DS$_A$, DS$_B$, DS$_D$, h$_D$)
5) B → A    RREP    (S, D, Seq, <S, A, B, D>, DSs, DS$_A$, DS$_B$, DS$_D$, h$_D$)
6) A → S    RREP    (S, D, Seq, <S, A, B, D>, DSs, DS$_A$, DS$_B$, DS$_D$, h$_D$)

**Description**

Source S initiates route discovery process by generating route request (RREQ) message. On every broadcast of RREQ, life time of RREQ would be decreased by one. RREQ would be discarded if life time reached to zero. Source appends its Digital Signature (DS$_S$) and hash code h$_S$ = H(S, D, Seq, L$_{RREQ}$).

When a neighbor of S, say A receives the RREQ message, it verifies the signature of S. and appends its identifier A to route list and its Digital Signature (DS$_A$) to RREQ and replaces h$_S$ by h$_A$= H(h$_S$, A, L$_{RREQ}$-1).

Similarly, next node B verifies signatures of node(s) of route list and then appends its identifier B and Digital Signature (DS$_B$) to RREQ, and replaces h$_A$ by h$_B$= H (h$_A$, B, L$_{RREQ}$-2).

Finally, when destination D receives the RREQ, it verifies all the signatures; it computes hash code h$_D$ to check integrity of RREQ:

$$h_D = H(B, L_{RREQ}\text{-}2\ H(A, L_{RREQ}\text{-}1\ H(S, D, Seq, L_{RREQ})\ ))$$

This must be same as h$_B$. If both values are same, D sends back route reply (RREP) message to B towards S. Otherwise discards RREQ. When S receives RREP, S verifies signature of all nodes S also computes hash code h to check message integrity:

$$h = H (B, L_{RREQ}\text{-}2\ H (A, L_{RREQ}\text{-}1\ H (S, D, Seq, L_{REQ})))$$

h must be same as h$_D$. Otherwise RREP is discarded by S.

## 4.2 Inter Group Communication

Inter group communication are done via group leaders. If a node receives RREQ from other's group node, RREQ would be discarded. Verification of nodes and message integrity is maintained as in intra group communication using digital signature and hash code of message; but when destination node is not member of same group as source nodes' group, this RREQ is forwarded to group leader. Group leader send route request RREQ to destination's group leader. Group leader send RREQ message to destination node maintaining a route from itself to destination. Finally, destination node replies back to its group leader by sending route reply RREP message towards source node.

## 5. SECURITY ANALYSIS OF PROPOSED SOLUTION

In this section, we discussed the security analysis of proposed key management system and routing protocol against different attacks.

### 5.1 Key Management Scheme

#### 5.1.1 Backward Secrecy

When a node leaves the network, it should not be able decrypt the future encrypted traffic. In proposed key management scheme, whenever a node leaves the group, group leader regenerates new group key and distribute it in the group. On the other hand, when a group leader leaves the network, a new group leader generates group key for the group. This ensures that keys are updated and backward secrecy is maintained in network.

### 5.1.2 Forward Secrecy

Forward secrecy says that when a new node joins the network, it should not be able to decrypt the past encrypted traffic. On joining of new node, group leader generates new group key and sends to members of group encrypted with old group key and unicasts to new node encrypted with key shared between group leader and new node, ensuring forward secrecy.

### 5.1.3 Mutual Authentication

In proposed key management system, both new node and group leader authenticate each other mutually at the time of network joining. After successful mutual authentication, node can join the network. When two nodes wish to communicate, they also authenticate each other by sending their Digital Signature.

### 5.1.4 Man in Middle Attack

Man in the Middle (MITM) attack is a kind of active attack in which an attacker remains invisible between two nodes say A and B. Attacker splits the connection into two connections, one between node A and attacker and second, between attacker and second node B. Two nodes A and B think that they are communicating with each other, while they communicate with attacker seating in between them. Key management system proposed in [12] is vulnerable to MITM attack; where an initiator (new joining node) sends its public key to central node. In the response of request, central node generates a session key and sends to initiator, encrypted with initiator's public key. In this scheme, an attacker may exist in between initiator and central node; attacker can capture the public key of new node and send its public key to central node. Then central node shares the session key with attacker and attacker shares session key with initiator. But in proposed key management system, both new node and group leader authenticate each other using challenge-response protocol. Hence, our key management system is not vulnerable to MITM attack.

## 5.2 Routing Protocol

In proposed routing protocol route request message RREQ carries hash code of parameters contained in RREQ. Therefore, any kind of modification in RREQ would change hash code. RREQ carries a destination sequence number preventing reply attack and to avoid loop. RREQ also carries Digital Signature of all nodes in route from source to destination. RREQ packet will not go in infinite loop, because life time message decreases by 1 on every broadcast.

Proposed routing protocol also provides security against MITM attack. Where an attacker X keeps itself invisible and makes changes in message without showing itself. Suppose, in above operation an attacker X is present in between node A and B and $L_{RREQ}=8$, then

1)     S → *    RREQ   (S, D, Seq, 8, < S>, DSs, $h_S$)
$$h_S = H(S, D, Seq, 8)$$
2)     A → *    RREQ   (S, D, Seq, 7, <S, A>, DSs, $DS_A$, $h_A$)
$$h_A = H(h_S, A, 7)$$
3)     X → *    RREQ   (S, D, Seq, 6, <S, A>, DSs, $DS_A$, $h_A$)

73

Attacker X would not do any kind of modification in RREQ, but life time of RREQ is decreased by one automatically. Therefore, node B receives RREQ (S, D, Seq, 5, <S, A>, DSs, $DS_A$, $h_A$), instead of RREQ (S, D, Seq, 6, <S, A>, DSs, $DS_A$, $h_A$). B calculates $h_B$ = H($h_A$, B, 5) and rebroadcast RREQ message. When destination D receives RREQ (S, D, Seq, 4, <S, A, B>, DSs, $DS_A$, $DS_B$, $h_B$), D computes $h_D$ = H (B, 5, H (A, 6, H (S, D, Seq, 7))) which is not same as $h_B$. So, D will discard the RREQ message. Hence, our protocol is not vulnerable to MITM attack.

## 6. CONCLUSION

In this paper, we proposed a key management scheme and a secure routing protocol for mobile ad hoc networks. We described a secure key management system for group based a mobile ad hoc network that does not rely on a centralized authority for generating and distributing keys. Group leaders generate, maintain, and distribute the keys in their groups in a secure manner. Challenge-response protocol allows a new incoming node to authenticate to group leader, then joins group. Proposed routing protocol uses hash function to maintain the integrity of message. Therefore, any kind of modification in RREQ can be detected. Using Public Key Cryptography (PKC), nodes can negotiate the session key for secure communication that fulfills the requirement of confidentiality. Security analysis results show that protocol establishes a route secure from different kind of attacks such as reply attack, rushing attack, IP spoofing and man in the middle attack.

Proposed key management is a decentralized and hybrid scheme combining both symmetric and asymmetric cryptographic algorithms; which maintains forward and backward secrecy and provides security against many attacks such as reply attack, man in the middle attack etc. Limitation of proposed key management system and routing protocol is that both use public key cryptography for key sharing and digital signature, which consumes more battery power in comparison of symmetric key cryptography.

## REFERECES

[1] L. Zhou and Z. J. Haas, "Securing ad hoc networks," IEEE Network, vol. 13, no. 6, pp: 24–30, 1999.
[2] Meng Ge, Kwok-yan Lam, "Self-healing Key Management Service for Mobile Ad Hoc Networks", Proceeding of first International Conference on Ubiquitous and Future Networks", June, 2009.
[3] S. Yi and R. Kravets, "MOCA: Mobile Certificate Authority for Wireless Ad hoc networks," 2nd Annual PKI Research Workshop (PKI 03), 2003.
[4] H. Y. Luo, J. J. Kong, P. Zerfos, S. W. Lu, and L. X. Zhang, "Ursa: Ubiquitous and robust access control for mobile ad hoc networks, " IEEE/ACM Transactions on Networking, vol. 12, no. 6, pp: 1049–1063, 2004.
[5] Mhd. Al-Shurman, Seong-Moo, Yoo, Bonam Kim, "Distributive Key Management for Mobile Ad Hoc Networks", International Conference on Multimedia and Ubiquitous Engineering, pp: 533-536, 2008.
[6] N. Kettaf, H. Abouaissa, P. Lorenz, "An Efficient Heterogeneous Key Management approach For Secure Multicast Communication in Ad hoc networks", Springer, Telecommunication System, vol-37, pp: 29-36 , February 2008.
[7] R. Blom, "Optimal class of symmetric key generation systems", Proceeding of the EUROCRYPT 84 workshop on Advances in Cryptology: Theory and Application of Cryptographic Techniques, pp: 335-338, December 1985, Paris, France.
[8] H. Nam Nguyen, H. Morino, "A Key Management Scheme for Mobile Ad hoc Networks Based on Threshold Cryptography for Providing Fast Authentication and Low Signaling Load", EUC Workshops-2005, LNCS 3823, pp: 905-915, 2005.

[9] Zhu Lina, Zhang Yi, Feng Li, "Distributed Key Management in Ad hoc Network based on Mobile Agent", Proceeding of 2nd IEEE International Symposium on Intelligent Information Technology Application, vol. 1, pp: 600-604, 2008.

[10] G. A. Safdar, C. McGrath, M. McLoone, "Limitations of Existing Wireless Networks Authentication and Key Management Techniques for MANETs", Proceeding of 7th IEEE International Symposium on Computer Networks, pp: 101-107, 2006.

[11] Yang Ya-Tao, Zeng Ping, and Fang Yong, Chi Ya-Ping., "A Feasible Key Management Scheme in Ad hoc Network", Proceeding of 8th IEEE International Conference on Software Engineering, Artificial Intelligence, Networking and Parallel/Distributed Computing (SNPD), pp: 300–303, 2007.

[12] Azzendine Boukerche and Yonglin Ren, "The Design of a Secure Key Management System for Mobile Ad Hoc Networks", The 33rd IEEE Conference on Local Computer Networks, pp: 302-327, October, 2008.

[13] Xukai Zou, Byrav Ramamurthy, "A Simple Group Diffie-Hellman Key Agreement Protocol without Member Serialization", Computational and Information Science, LNCS-3314, pp: 725-731, 2004.

[14] W. Huang, Y. Xiong, and D. Chen, "DAAODV: A Secure Ad hoc Routing Protocol based on Direct Anonymous Attestation", Proceeding of International Conference on Computational Science and Engineering, August, vol-2, pp: 809-816, 2009.

[15] D. Cerri and A. Ghioni, "Securing AODV: The A-SAODV Securing Routing Prototype", IEEE Communication Magazine: Security in Mobile Ad hoc and Sensor Networks, vol-46, pp: 120-125, February, 2008.

[16] P. Papadimitratos, and Z. Haas, "Secure Routing for Mobile Ad hoc Networks", Proceeding of SCS Communication Networks and Distributed Systems Modeling and Simulation, January, 2002.

[17] Y.C. Hu, A. Perrig, and D. B. Johnson, "Ariadne: A Secure On-demand Routing Protocol for Ad hoc Networks", Proceeding of 8th Annual International Conference on Mobile Computing and Networking, (MobiCom 02), September 2002, pp: 12-23,.

[18] J. Liu, F. Fu, J. Xiao and Y. Lu, "Secure Routing for Mobile Ad Hoc Networks", Proceeding of 8th ACIS International Conference on Software Engineering, Artificial Intelligence, Networking, and Parallel/Distributed Computing, vol-3, 2007, pp: 314-318.

[19] L. Buttyan, and I. Vajda, "Towards Provable Security for Ad hoc Routing Protocols", Proceeding of 2nd ACM Workshop on Security of Ad hoc and Sensor Networks, October 2005, pp: 94-105.

[20] G. Ács, L. Buttyán, and I. Vajda, "Provably Secure On-demand Source Routing in Mobile Ad Hoc Networks", IEEE Transactions on Mobile Computing, vol-5, November 2006, pp: 1533-1546.

[21] N. Kettaf, H. Abouaissa, P. Lorenz, "An Efficient Heterogeneous Key Management approach For Secure Multicast Communication in Ad hoc networks", Springer, Telecommunication System, vol-37, February 2008, pp: 29-36.

[22] Y.Chun Hu, A. Perrig and David B. Johnson, "Wormhole Attack in Wireless Networks", IEEE Journal on Selected Areas in Communication, vol. 24, February 2006, pp: 370-380.

[23] R.A. Raja Mahmood, A.I. Khan, "A Survey on Detecting Black Hole Attack in AODV-Based Mobile Ad hoc Networks", International Symposium on High Capacity Optical Networks and Enabling Technologies, November 2007, pp: 1-6.

[24] A. K. Shukla, N. Tyagi, "A New Route Maintenance in Dynamic Source Routing Protocol", IEEE International Symposium on Wireless Pervasive Computing, January, 2006.

[25] Q. Niu, "Secure On-Demand Source Routing for Ad hoc Networks", Proceeding of IEEE International Conference on Wireless Communications, Networking and Mobile Computing (WiCOM 08), October, 2008, pp: 1-4.

[26] Du Congwei, Li Rongsen and Dou Wenhua, "An Efficient Key Agreement Protocol in Cluster-Based MANETs", IEEE International Conference on Computer Application and System Modeling, Taiyuan, China, October 22-24, 2010, vol-10, pp: v10-627-v10-630.

[27] Xie Hai-tao, "A Cluster-Based Key Management Scheme for MANET", Proceeding of IEEE 3rd International Workshop on Intelligent System and Application, May 28-29, 2011, Wuhan, China, pp:1-4.